\documentclass[12pt,a4paper]{article}

\usepackage[truedimen,margin=30mm]{geometry} 

\usepackage{amssymb}
\usepackage{amsmath}
\usepackage[hiresbb]{graphicx}
\usepackage{amsthm}   
\usepackage{authblk}   
\usepackage{lscape}   
\usepackage{url}
\usepackage{multirow}
\usepackage{natbib}

\usepackage{setspace}

\usepackage{times}

\makeatletter

\@addtoreset{equation}{section}
\makeatother

\makeatletter
\def\mojiparline#1{
    \newcounter{mpl}
    \setcounter{mpl}{#1}
    \@tempdima=\linewidth
    \advance\@tempdima by-\value{mpl}zw
    \addtocounter{mpl}{-1}
    \divide\@tempdima by \value{mpl}
    \advance\kanjiskip by\@tempdima
    \advance\parindent by\@tempdima
}
\makeatother
\def\linesparpage#1{
    \baselineskip=\textheight
    \divide\baselineskip by #1
}

\allowdisplaybreaks[4]

\title{ {\bf Efficient Prior Sensitivity and Tipping-point Analysis for Medical Research: Revisiting Sampling Importance Resampling} }

\author[1]{Tomohiro Ohigashi}
\author[2]{Shonosuke Sugasawa}

\affil[1]{Department of Information and Computer Technology, Faculty of Engineering, Tokyo University of Science, Tokyo, Japan}
\affil[2]{Faculty of Economics, Keio University, Tokyo, Japan}

\date{}

\begin{document}
\linesparpage{25}
\allowdisplaybreaks[4]
\begin{singlespace}
\maketitle
\end{singlespace}

\vspace{-1cm}
\begin{center}
{\large\bf Abstract}
\end{center}
Bayesian  methods have received increasing attention in medical research, where sensitivity analysis of prior distributions is essential.
Such analyses typically require the evaluation of the posterior distribution of a parameter under multiple alternative prior settings. 
When the posterior distribution of the parameter of interest cannot be derived analytically, the standard approach is to re-fit the model using Markov chain
Monte Carlo (MCMC) for each setting, which incurs substantial computational costs.
This issue is particularly relevant in tipping-point analysis, in which the posterior must be evaluated across gradually changing degrees of borrowing.
Sampling-importance resampling (SIR) provides an efficient alternative by approximating posterior samples under new settings without MCMC re-fitting.
Despite its potential computational advantages, the practical performance of SIR in repeated prior-sensitivity analyses, including tipping-point analysis, and in complex Bayesian models used in medical research has not been sufficiently illustrated.
In this study, we illustrate the practical utility of SIR through two case studies: one involving tipping-point analysis under external data borrowing and another involving sensitivity analysis for a Bayesian nonparametric model in meta-analysis.
In both examples, SIR substantially reduced computation time and produced posterior summaries similar to those obtained by MCMC re-fitting.


\section{Introduction}
In recent years, Bayesian methods have attracted considerable attention in medical research because they can easily measure uncertainty and incorporate prior information \citep{goligherBayesianStatisticsClinical2024}.
In the field of clinical trials, Bayesian methods have been increasingly applied in early-phase development, such as phase I oncology trials.
More recently, their use has also begun to be explored  in later-phase development \citep{ bswg}.
The Complex Innovative Trial Design (CID) pilot meeting program, which began in 2018, gives applicants and regulatory authorities a chance to discuss complex trial designs and analytical methods, including Bayesian methods \citep{priceUSFoodDrug2021}.
Several trials that joined the CID pilot meeting program have been reported, some of which included Bayesian methods in their statistical analysis plan.
Since 2023, the CID pilot meeting program has been operating  as the CID paired meeting program \citep{FDA_CID_MeetingProgram}.
Thus, Bayesian methods are expected to continue to be used in clinical trials.

In Bayesian analysis, sensitivity analysis of the prior distributions is generally considered essential.
The U.S. Food and Drug Administration’s guidance on Bayesian analysis in medical device clinical trials also recommends submitting sensitivity analyses \citep{foodanddrugadministrationfdaGuidanceUseBayesian2010}. 
The guidance requires that sensitivity analyses be performed with respect to model assumptions, prior distributions, and hyperparameters.
These  analyses require derivation of the posterior distribution of the parameter of interest for each alternative setting.
When the posterior distribution of the parameter of interest cannot be derived analytically, the standard approach is to re-fit the Markov chain Monte Carlo (MCMC) algorithm for each setting.
The ICH E11A guideline ``Pediatric Extrapolation,'' finalized in 2024, references tipping-point analyses of parameters that determine the degree of borrowing from adult data when such information is incorporated into the prior distribution for pediatric trial analysis \citep{ICH_E11A_Step4_2024_0821}. 
In tipping-point analysis, the posterior distributions of the parameter of interest should  be obtained while gradually changing the degree of borrowing.
If MCMC is required to obtain the posterior distributions, the model must be repeatedly re-fitted using MCMC, resulting in a high computational cost.
Recent advances in software for implementing MCMC algorithms, such as Stan and JAGS, have reduced the computational  burden associated with fitting models using MCMC \citep{ carpenterStanProbabilisticProgramming2017, plummerJAGSProgramAnalysis2003}.
Nevertheless, fitting complex statistical models can still be computationally intensive, even for a single MCMC run.
Accordingly, in situations such as tipping-point analysis, where repeated MCMC runs are required, the computational cost remains a considerable challenge.

Sampling-importance resampling (SIR) is a method for obtaining samples from the posterior distribution under an alternative setting without MCMC re-fitting
\citep{rubinCalculationPosteriorDistributions1987,smithBayesianStatisticsTears1992}.
The utility of the SIR algorithm has been recognized when computational cost is a major limitation, and it is commonly presented in textbooks on Bayesian statistics
\citep{gelmanBayesianDataAnalysis2013,lesaffre2012bayesian}.
Recent work has also shown that posterior expectations and divergence-based sensitivity measures under alternative priors can be evaluated by Monte Carlo integration using draws from a single base posterior, without model re-fitting
\citep{sugasawaPriorSensitivityAnalysis2026}.
However, the practical performance of SIR-based posterior sampling has not been sufficiently illustrated in settings that require repeated prior-sensitivity analyses, such as tipping-point analysis, or in applications involving complex Bayesian models used in medical research.
In this study, we illustrate the practical utility of the SIR algorithm through two case studies.
The first case considers tipping-point analysis under external data borrowing, whereas the second considers prior-sensitivity analysis for a Bayesian nonparametric model in meta-analysis.

The remainder of this paper is organized as follows. 
In Section 2, we introduce the SIR algorithm, the effective sample size as a diagnostic of importance-weight degeneracy, and a grid-based formulation of tipping-point analysis.
In Section 3, we conduct two case studies. 
The first examines a tipping-point analysis in a setting involving external data borrowing, while the second addresses a sensitivity analysis of a nonparametric Bayesian model applied to meta-analysis.
We conclude our paper in Section 4 with further discussion.

\noindent

\section{Method}

\subsection{Sampling importance resampling algorithm}

Suppose that we are interested in fitting a model $f(x \mid \theta)$ to observed data $x$, where $\theta$ is an unknown parameter. 
To perform Bayesian inference on $\theta$, we assign a prior distribution $\pi(\theta)$ on $\theta$ and consider a posterior distribution $\pi(\theta \mid x)\propto \pi(\theta)f(x\mid\theta)$. 
Suppose we have a ``base'' posterior $\pi(\theta \mid x) \ \propto\ \pi(\theta)\, f(x \mid \theta)$, and we wish to conduct inference under an ``alternative'' prior $\pi_\ast(\theta)$, namely, $\pi_\ast(\theta \mid x) \ \propto\ \pi_\ast(\theta) f(x \mid \theta)$. 
Let $\{\theta_{(m)}\}_{m=1}^M$ be posterior draws from $\pi(\theta \mid x)$. 
As  the likelihood is unchanged, the ratio of the two posteriors reduces to the prior ratio, and the normalized importance weights are given by
\begin{equation*}
 \tilde w_m = \frac{w_m}{\sum_{j=1}^M w_j}, \ \ \ \ \ 
 w_m \ \propto\ \frac{\pi_\ast(\theta_{(m)})}{\pi(\theta_{(m)})}.
\end{equation*}
Then, for any integrable functional $h(\theta)$, the posterior expectation under the alternative posterior can be approximated as $\mathbb{E}_{\pi_\ast(\theta \mid x)}[h(\theta)]\approx \sum_{m=1}^M \tilde w_m h(\theta_{(m)})$.
Credible intervals can be obtained either from the weighted empirical distribution of $\{\theta_{(m)}\}$ or by a SIR  step that resamples 
$\theta^\dagger_{(1)}, \ldots, \theta^\dagger_{(M^\dagger)}$ 
from $\{\theta_{(m)},\  m=1,\ldots,M\}$ with probabilities $\{\tilde w_m, \ m=1,\ldots,M\}$ \citep{rubinCalculationPosteriorDistributions1987,smithBayesianStatisticsTears1992}.

SIR is reliable only when the support of $\pi_\ast(\theta)$ is contained within the support of $\pi(\theta)$ and when the weight distribution is not overly heavy-tailed \citep{kong1994sequential,liu2001monte}. 
A common diagnostic is the effective sample size (ESS), defined as ${\rm ESS} = 1/\sum_{m=1}^M \tilde w_m^2$ with a small ESS indicating unreliable reweighting \citep{kong1994sequential,liu2001monte}. 
Using SIR, posterior summaries under an alternative prior can be obtained without re-fitting the model. 
By repeating this procedure for a range of alternative priors, we can examine how posterior quantities such as the mean, variance, and credible intervals change across different prior specifications. 
This enables a straightforward method of prior sensitivity analysis, allowing us to assess the robustness of the posterior inference for the choice of prior distribution \citep[for example, ][]{berger2000bayesian,roos2015sensitivity}.

\subsection{Tipping-point analysis}

Tipping-point analysis evaluates how the posterior conclusion changes as a prior hyperparameter is varied. Suppose that $\psi$ denotes a scalar hyperparameter governing the prior distribution and that $\theta_0$ is a prespecified null value for the parameter of interest. For each candidate value of $\psi$, posterior summaries can be obtained by reweighting posterior draws generated under a base prior, without re-fitting the model.

In practice, we evaluate an ordered grid
$\Psi=(\psi_1,\ldots,\psi_K)$, where $\psi_1$ denotes the
base-prior specification and the remaining values are arranged according
to a prespecified search direction.
The grid-based tipping point is defined as the first grid value encountered
along this sequence at which a specified posterior decision criterion is met.
For example, when treatment benefit is concluded if the upper limit of a
two-sided $100(1-\alpha)\%$ credible interval is below $\theta_0$,
we define
\[
k_{\mathrm{tip}}
=
\min\left\{
k\in\{1,\ldots,K\}:
U_{1-\alpha/2}(\psi_k)<\theta_0
\right\},
\qquad
\psi_{\mathrm{tip}}=\psi_{k_{\mathrm{tip}}},
\]
where $U_{1-\alpha/2}(\psi_k)$ denotes the
$(1-\alpha/2)$-quantile of the posterior distribution under
hyperparameter value $\psi_k$.

\section{Applications}

\subsection{Tipping-point analysis for incorporating external data}

A phase III randomized controlled trial (RCT) for first-line diffuse large B-cell lymphoma, hereafter referred to as the DLBCL trial, is included in the Complex Innovative Trial Design pilot meeting program \citep{fdadlbcl}.
The primary endpoint of the DLBCL trial is progression-free survival (PFS), and it is analyzed based only on the RCT data.
The key secondary endpoint is overall survival (OS).
OS is analyzed at the time when the number of PFS events reaches the planned number of events.
Consequently, the statistical power for OS is expected to be insufficient.
Hence, patients are selected from an external data source to form an external control group, which is incorporated into the OS analysis through a commensurate prior \citep{hobbsCommensuratePriorsIncorporating2012}.
In the commensurate prior, the degree of borrowing from external control data is adjusted based on the conflict between the external and  current control data.
The choice of hyperparameters for the commensurate prior may affect the conclusions regarding the treatment effect, depending on the current data.
Therefore, this case study focuses on tipping-point analysis for OS analysis in the DLBCL trial.

\begin{figure}
    \centering
    \includegraphics[width=0.5\linewidth]{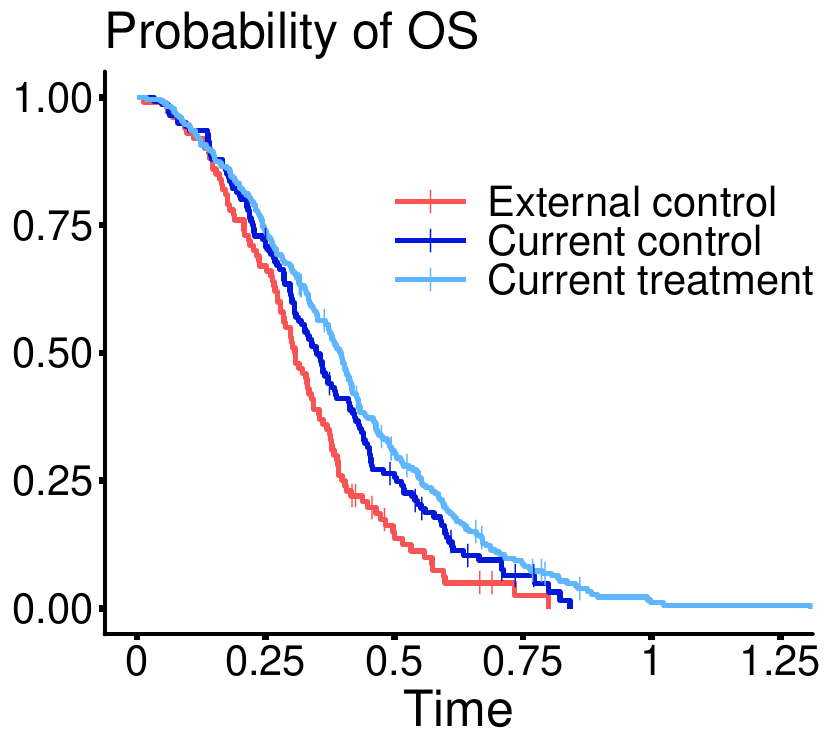}
    \caption{Kaplan--Meier plot for the case study}
    \label{km}
\end{figure}

The settings for the single simulated dataset were determined in accordance with materials provided by the Bayesian Scientific Working Group \citep{bswg}.
The sample sizes were set to 280 for the current treatment group, 140 for the current control group, and 100 for the external control group.
For simplicity, this case study treats the external control data as already selected from an external data source and fixed.
Figure \ref{km} shows a Kaplan--Meier plot based on the simulated dataset.
As the aim of this case study is to explore the computational efficiency of tipping-point analysis using the SIR algorithm, the simulated data were generated such that the number of events was sufficient and the data showed evidence of treatment benefit, in contrast to the settings of the DLBCL trial.
We adopted a proportional hazards model for OS.
We denoted the time-to-event for patient $i$ by $t_i$, the log-hazard ratio by $\beta$, and the treatment group indicator by $z_i$.
For patient $i$ in the current treatment and control groups, the hazard function was assumed to be $h_i(t)=h_0(t)\exp(\alpha_{\mathrm{C}}+\beta z_i)$, where $\alpha_{\mathrm{C}}$ denotes the log-hazard parameter for participants in the current trial.
For patient $i$ in the external control group, the hazard function was assumed to be $h_i(t)=h_0(t)\exp(\alpha_{\mathrm{E}})$, where $\alpha_{\mathrm{E}}$ denotes the log-hazard parameter for participants in the external control group.
We assumed a Weibull model for the baseline event-time distribution, so that $h_0(t)$ had the corresponding Weibull hazard form.
The Weibull shape parameter was assigned an $\operatorname{Exp}(0.001)$ prior.
We assigned vague $\mathrm{N}(0,10^2)$ priors to $\beta$ and $\alpha_{\mathrm{E}}$.
For $\alpha_{\mathrm{C}}$, we assigned the commensurate prior
$\mathrm{N}(\alpha_{\mathrm{E}},\tau^2)$.
The choice of the hyperprior for $\tau$ is important because it is related to the degree of borrowing from the external control data.
Therefore, we conducted a tipping-point analysis by assigning a half-normal prior $\operatorname{HalfNormal}(s)$ to $\tau$, where $s$ denotes the scale parameter, and examining the posterior distribution of the hazard ratio while varying $s$.

To evaluate the computational efficiency achieved by the SIR algorithm, we considered 91 candidate values, $s=0.10,0.11,\ldots,1.00$.
For the tipping-point analysis, these values were examined from the base-prior value $s=1.00$ downward in increments of 0.01.
As a comparator, we re-fitted the model using MCMC for each value of $s$.
In the SIR approach, $s=1.00$ was used as the base-prior specification because it corresponds to the most diffuse half-normal prior among the candidate specifications and is therefore expected to provide adequate overlap with the alternative priors.
For each alternative value, $s=0.10,0.11,\ldots,0.99$, posterior samples of the hazard ratio were obtained using the SIR algorithm.
Posterior sampling was performed using the No-U-Turn Sampler (NUTS), an adaptive Hamiltonian Monte Carlo algorithm, implemented in the \texttt{cmdstanr} version 0.4.0 package in R version 4.4.3.
We ran four parallel MCMC chains, each with 5,000 warm-up iterations followed by 100,000 sampling iterations, retaining every fifth post-warm-up draw. The target acceptance probability was set to 0.9999 to reduce divergent transitions by using a smaller integration step size.
The same sampling configuration was used for each MCMC re-fit and for the single MCMC fit under the base prior in the SIR approach.
For each alternative prior specification, we generated an equally weighted posterior sample by multinomial resampling according to the normalized importance weights. 
The resample size was set to $M^\dagger = 0.8 M$, where $M$ denotes the number of posterior draws obtained under the base prior.

\begin{figure}[htb!]
\centering
\includegraphics[width=1\linewidth]{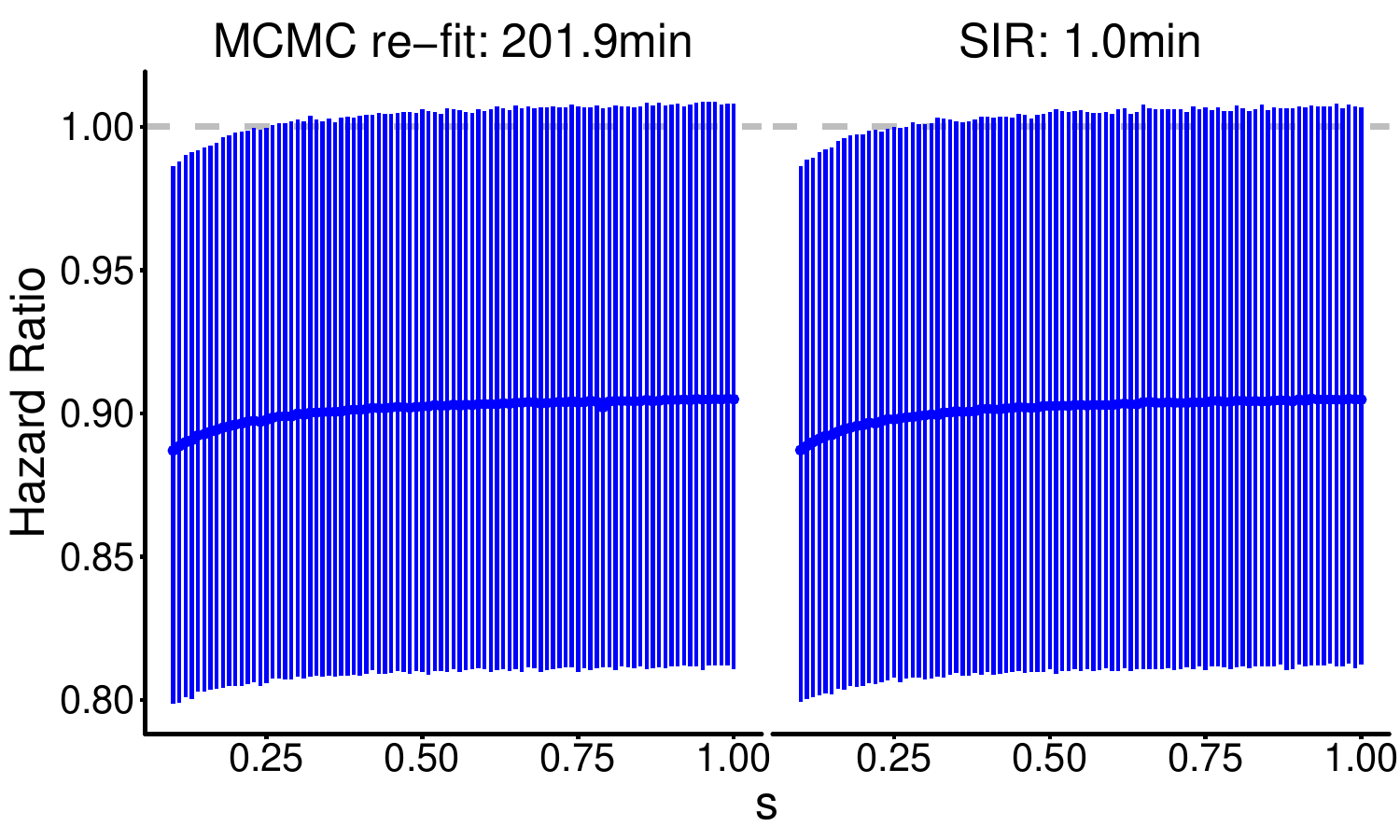}
\caption{Tipping-point analysis for the MCMC re-fit (left) and SIR (right) approaches}
\label{tipping1}
\end{figure}

\begin{figure}[htb!]
    \centering
    \includegraphics[width=0.5\linewidth]{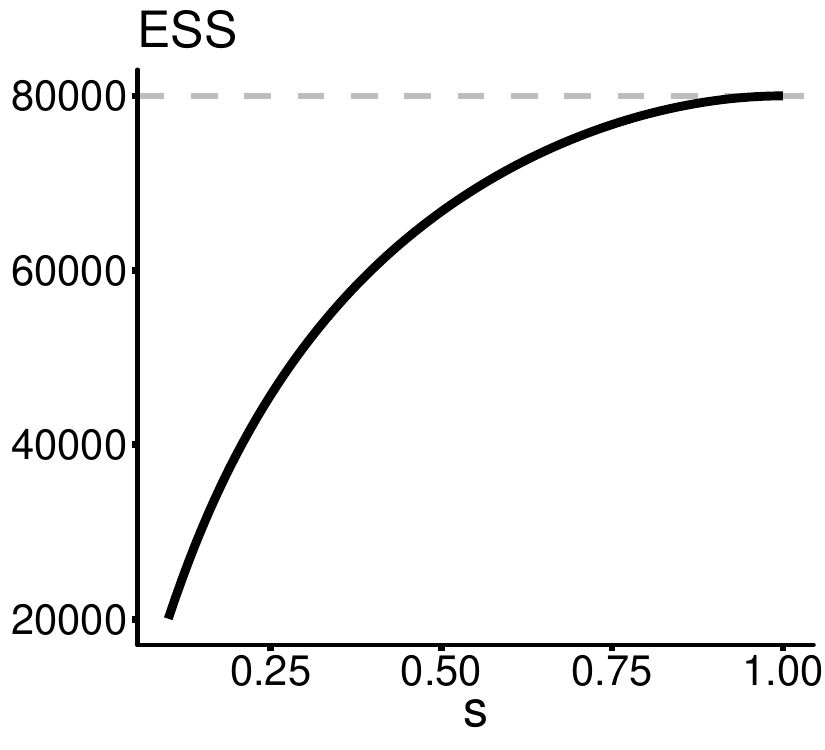}
    \caption{Effective sample size for the SIR approach}
    \label{ess1}
\end{figure}

Figure \ref{tipping1} shows the posterior mean and 95\% credible interval of the hazard ratio as a function of the hyperparameter $s$ for both the MCMC re-fit and SIR approaches.
Overall, no substantial differences were observed between the two approaches.
We evaluated 91 candidate values of the half-normal scale parameter, $s=0.10,0.11,\ldots,1.00$. 
The grid-based tipping point was defined as the first evaluated value encountered when moving downward from the base-prior value $s=1.00$ for which the upper limit of the 95\% posterior credible interval for the hazard ratio was below 1.0.
The resulting tipping points were 0.25 for the MCMC re-fit approach and 0.26 for the SIR approach.
Computation times were 201.9 minutes for the complete set of 91 MCMC re-fits and 1.0 minute for the SIR approach, including the single base-prior MCMC fit and the subsequent reweighting and resampling over all candidate values.
Figure \ref{ess1} shows the ESS calculated from the normalized importance weights before resampling as a function of the hyperparameter $s$.
The ESS was 80,000 under the base prior ($s=1.0$) and decreased as $s$ decreased.
This decline was expected because the overlap between the base and alternative priors decreased as $s$ moved away from the base-prior value. 
At the lowest value of $s=0.1$, the ESS was 20,000.

\subsection{Sensitivity analysis for nonparametric Bayesian meta-analysis}
In meta-analyses for studies with substantial variability in quality, it is necessary to address internal validity biases.
As internal validity biases are not directly observable, correcting for them in a meta-analysis remains a challenging task.
To address this issue, \citet{verdeBiasCorrectedBayesianNonparametric2025} proposed a bias-corrected Bayesian nonparametric (BC-BNP) model.
The BC-BNP model introduces an indicator variable $I_i$ that indicates
whether study $i$ is biased.
It assumes that each study has the same probability of being biased.
This probability, denoted by $\pi^{\mathrm{B}}$, represents the probability that a study is biased.
Although \citet{verdeBiasCorrectedBayesianNonparametric2025} proposed a procedure to determine the hyperparameters of the prior for $\pi^\text{B}$, a sensitivity analysis for this prior remains necessary.
Therefore, this case study focuses on a sensitivity analysis of the prior for $\pi^\text{B}$.

Suppose that a meta-analysis of $N$ studies reports effect estimates $y_1, y_2, \dots, y_N$ with their corresponding standard errors $\text{SE}_1, \text{SE}_2, \dots, \text{SE}_N$. 
If the outcome for each study is binary, $y_i$ represents the log-odds ratio.
\citet{verdeBiasCorrectedBayesianNonparametric2025} assumed that the reported effect estimate $y_i$
follows
\[
y_i \mid \theta_i^{\text{B}}
\sim
\mathrm{N}\left(\theta_i^{\text{B}},\mathrm{SE}_i^2\right),
\qquad
\theta_i^{\text{B}}=\theta_i+I_i\beta_i,
\]
where $\theta_i$ denotes the bias-corrected study effect,
$\beta_i$ represents the internal validity bias for study $i$,
and $I_i$ indicates whether study $i$ is biased.
The random effect $\theta_i$ is modeled as $\theta_i \sim \text{N} (\mu_\theta, \tau^2_\theta)$, where $\mu_\theta$ represents the mean effect and $\tau^2_\theta$ represents the between-study variance among the bias-corrected studies.
A Dirichlet process (DP) prior is assigned to $\beta_i$, having a base distribution $\text{N}(\mu_\beta, \tau^2_\beta)$ and concentration parameter $\alpha$, described as   
\begin{align*}
\beta_i \mid G_\beta &\sim G_\beta, \ \ \ \ \ 
G_\beta \mid \mu_\beta, \tau^2_\beta, \alpha \sim \text{DP}\left( \text{N} (\mu_\beta, \tau^2_\beta), \alpha \right). 
\end{align*}
\citet{verdeBiasCorrectedBayesianNonparametric2025} employed a finite approximation with a maximum of $K$ components for the implementation of the DP.
Under the finite approximation of the Dirichlet process, the marginal distribution of the potentially biased study effect $\theta_i^{\mathrm{B}}$ is given by
\[
\theta_i^{\mathrm{B}}
\sim
(1-\pi^{\mathrm{B}})
\mathrm{N}(\mu_\theta,\tau_\theta^2)
+
\pi^{\mathrm{B}}
\sum_{k=1}^{K}
w_k^\star
\mathrm{N}(\mu_\theta+\beta_k^\star,\tau_\theta^2),
\]
where $w_k^\star$ denotes the stick-breaking weight,
$\beta_k^\star$ denotes the location of the $k$th bias component,
and $\pi^{\mathrm{B}}$ denotes the probability that a study is biased.
For the default prior for $\pi^\text{B}$, \citet{verdeBiasCorrectedBayesianNonparametric2025} assigned $\text{Beta}(0.5, 1.0)$.
The prior reflects the assumption that one-third of the studies in the meta-analysis are biased, that is, $\text{E}(\pi^\text{B}) = 1/3$.
In a case study on the relationship between hypertension and severity in COVID-19  patients---presented in \citet{verdeBiasCorrectedBayesianNonparametric2025}---the analysis was conducted by assigning an informative prior, $\text{Beta}(8.6, 1.97)$.
Although the default prior was used in their sensitivity analysis, the posterior distributions for the pooled odds ratio (OR)  differed between the informative and default priors.
In the present sensitivity analysis, we parameterize the prior distribution of $\pi^{\mathrm{B}}$ as $\pi^{\mathrm{B}} \sim \operatorname{Beta}(a_0,a_1)$, where $a_0>0$ and $a_1>0$ denote the two shape parameters.

To evaluate the computational efficiency achieved by the SIR algorithm,
we considered 54 candidate prior specifications comprising 18 values of $a_0$, ranging from 0.5 to 9.0 in increments of 0.5, and three values of $a_1$, namely, 1.0, 1.5, and 2.0.
As a comparator, we re-fitted the model using MCMC for each combination of $a_0$ and $a_1$.
In the SIR approach, the uniform prior $\pi^{\text{B}} \sim \mathrm{Beta}(1,1)$, corresponding to $a_0 = a_1 = 1.0$, was used as the base prior because it provides a relatively diffuse reference distribution over $(0,1)$.
Posterior samples of the pooled OR were then obtained for each alternative prior specification using the SIR algorithm.
Posterior sampling was performed using the \texttt{jarbes} package, which is built on JAGS.
For each MCMC re-fit, we ran four parallel chains with 220,000 iterations each, discarded the first 20,000 iterations as burn-in, and retained every fifth post-burn-in draw.
For the single MCMC fit under the base prior in the SIR approach, we ran four parallel chains with 320,000 iterations each, discarded the first 20,000 iterations as burn-in, and retained every fifth post-burn-in draw.
The remaining sampling settings were the same as those used for the MCMC re-fit approach.
For each alternative prior specification, we generated an equally weighted posterior sample by multinomial resampling according to the normalized importance weights.
The resample size was set to $M^\dagger=0.8M$, where $M$ denotes the number of posterior draws obtained under the base prior.

\begin{figure}[htb!]
    \centering
    \includegraphics[width=1\linewidth]{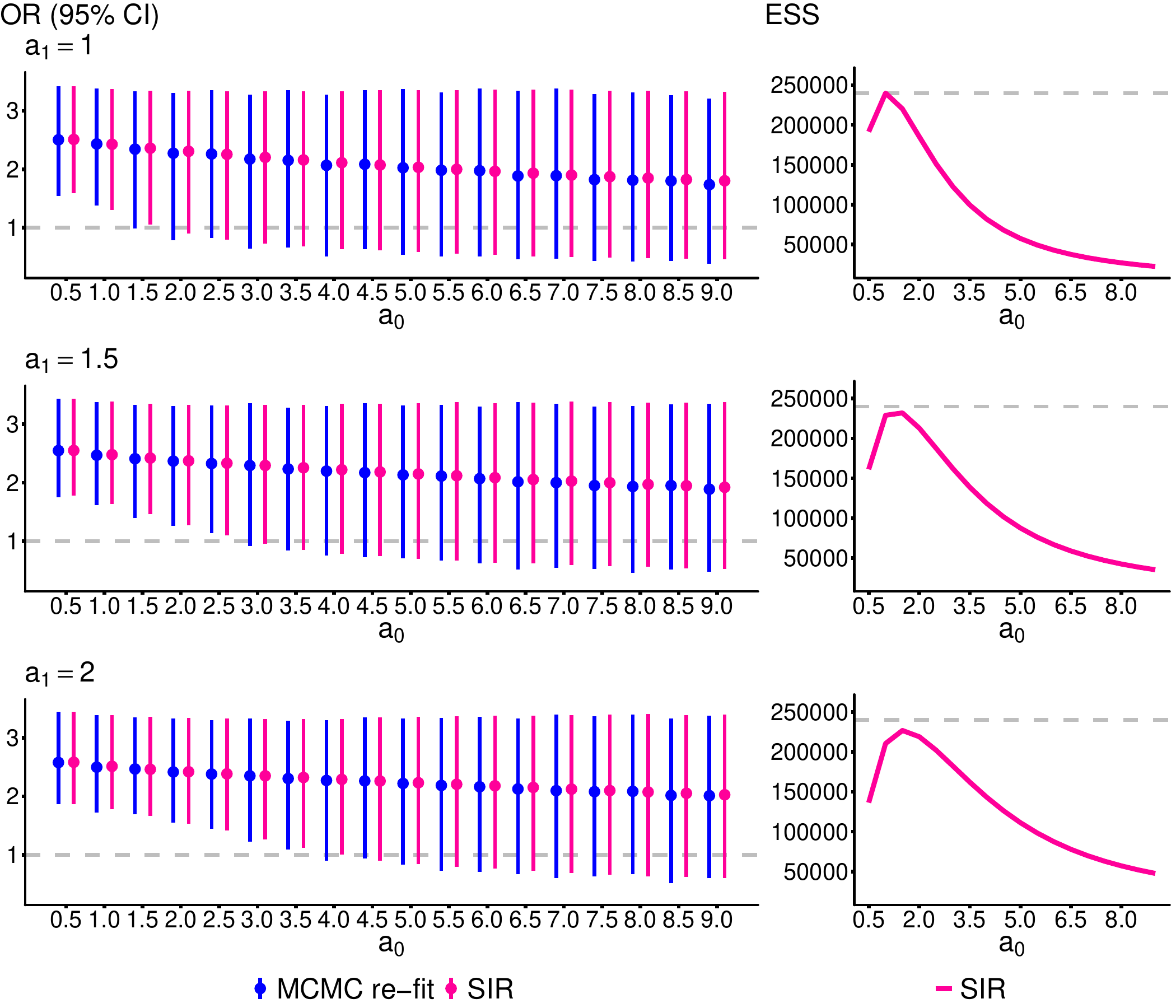}
    \caption{Sensitivity analysis for the MCMC re-fit and SIR approaches (left) and effective sample size for the SIR approach (right)}
    \label{case2}
\end{figure}

Figure \ref{case2} (left) shows the posterior mean and 95\% credible interval of the pooled OR as a function of the combination of $a_0$ and $a_1$ for both the MCMC re-fit and SIR approaches.
Overall, the posterior means and credible intervals obtained using the two approaches were similar. Minor differences in the interval limits were observed for some prior specifications, which may reflect Monte Carlo variability arising from both MCMC sampling and the resampling step.
Computation times were 295.6 minutes for the complete set of 54 MCMC re-fits and 8.6 minutes for the SIR approach, including the single base-prior MCMC fit and the subsequent reweighting and resampling over all candidate prior specifications.

Figure \ref{case2} (right) shows the ESS calculated from the normalized importance weights before resampling for the SIR approach.
The ESS was 240,000 under the base prior ($a_0=a_1=1.0$) and varied across the alternative prior specifications.
In general, the ESS tended to decrease as the alternative prior moved farther from the base prior, reflecting reduced overlap between the corresponding prior distributions.
The minimum ESS was 22,510 at $a_0=9.0$ and $a_1=1.0$.

\section{Discussion}
In communications between regulatory agencies and sponsors, such as the CID pilot/paired meeting program, it is often necessary to respond to regulatory requests within a limited timeframe. 
Therefore, improving the efficiency of the tasks handled by statisticians is critical.
In particular, when applying Bayesian methods using MCMC, reducing the time required for MCMC sampling can be highly beneficial.
In the context of clinical trial design, the SIR algorithm is particularly useful in simulation studies that aim to evaluate operating characteristics under various prior specifications.
For example, when assessing performance metrics through 10,000 simulation replicates, running MCMC separately for each prior setting would require 10,000 MCMC runs per specification, resulting in substantial computational costs.
By using the SIR algorithm, one can perform MCMC under a single baseline prior specification within each simulation replicate and then approximate the posterior samples corresponding to alternative prior specifications, thereby significantly reducing the computational burden required for the simulation study.

The use of the SIR algorithm is also valuable in post-hoc analyses, such as tipping-point analysis (e.g., Case 1), or in analyses conducted in response to peer-review or regulatory requests. In practice, one viable hybrid strategy would be to first use SIR to evaluate the posterior decision criterion over a prespecified grid of prior hyperparameter values. When the criterion changes monotonically with the hyperparameter, a standard root-finding procedure, such as bisection, could be used to refine the approximate tipping point within the interval identified by the grid search. If confirmation is required, the model could then be re-fitted using MCMC only at or near the estimated tipping point. This strategy could reduce computational cost while limiting the impact of approximation error on the final inference.

Several limitations should be noted. 
First, the reliability of SIR depends on sufficient overlap between the posterior distribution under the base prior and those under the alternative priors. 
In small-sample settings, the likelihood may provide limited information relative to the prior, so changes in the prior can induce larger shifts in the posterior distribution and may reduce the overlap between the base and alternative posteriors. 
Similarly, highly informative prior specifications can lead to poor posterior overlap when they differ markedly from the base prior. 
Where possible, the base prior should be chosen to be sufficiently diffuse to cover the posterior regions induced by the candidate priors, with the weight-based ESS examined separately for each alternative specification.
Although ESS provides a useful diagnostic of weight degeneracy, there is no universal ESS threshold that guarantees adequate approximation accuracy.

Separately, the accuracy of SIR depends on the number and quality of posterior draws obtained from the base-prior MCMC fit. 
Because the weight-based ESS is bounded above by the number of retained draws, shorter MCMC runs can lead to small ESS and unstable posterior summaries, particularly for tail quantiles. 
The base-prior chains should therefore be checked for convergence and adequate MCMC effective sample size before reweighting; if the weight-based ESS remains insufficient, the base run should be extended.

The present study was based on two case studies, and the approximation accuracy may differ across models, datasets, and prior specifications.
Accordingly, when the weight-based ESS is insufficient or when the posterior conclusion is close to a decision boundary, the SIR-based results should be confirmed by re-fitting the model under the relevant alternative prior using MCMC.

The SIR algorithm can also be applied in the presence of changes to the likelihood function by following a similar procedure.
For example, in Bayesian response-adaptive randomization \citep{robertsonResponseAdaptiveRandomizationClinical2023}, patients are sequentially allocated to treatment arms based on predictive probabilities conditioned on registered patient data. This design has been implemented in trials such as the BATTLE trials \citep{kimBATTLETrialPersonalizing2011, papadimitrakopoulouBATTLE2StudyBiomarkerIntegrated2016} and I-SPY 2 trial \citep{wangISPY2Neoadjuvant2019}, and its use is expected to become more widespread in the future.
However, when the likelihood is modified, the overlap between the pre- and post-change likelihood functions may be small, which can reduce the accuracy  of the importance weights.
This issue has been studied in the context of leave-one-out cross-validation and may be mitigated using Pareto smoothed importance sampling \citep{vehtariParetoSmoothedImportance2024}.

\section*{Acknowledgement}
This work was partially supported by JSPS KAKENHI Grant Numbers 24K20739, 24K21420, and  25H00546.

\section*{Data Availability Statement}
The data that support the findings of this study are openly available in the GitHub repository \url{https://github.com/tom-ohigashi/SIRforMedicalResearch}.

\bibliographystyle{chicago}
\bibliography{library}

\end{document}